\documentclass[12pt,english]{extarticle}
\usepackage[T1]{fontenc}
\usepackage[latin9]{inputenc}
\usepackage{geometry}
\usepackage{color}
\usepackage{babel}
\usepackage{amsmath}
\usepackage{amssymb}
\usepackage{graphicx}
\usepackage[unicode=true]
 {hyperref}

\makeatletter

\providecommand{\tabularnewline}{\\}

\newcommand{\lyxaddress}[1]{
	\par {\raggedright #1
	\vspace{1.4em}
	\noindent\par}
}

\date{}
\usepackage{lineno, setspace,hyperref} 
\hypersetup{
colorlinks=true,
linkcolor=red,
citecolor = blue,
urlcolor=blue,
pdfpagemode=FullScreen}

\usepackage{xcolor} 

\@ifundefined{showcaptionsetup}{}{%
 \PassOptionsToPackage{caption=false}{subfig}}
\usepackage{subfig}
\makeatother

\begin{document}
\title{Convexification of Charge Equilibrium within the Dendrites of Rechargeable
Batteries}
\author{Asghar Aryanfar$^{\dagger\ddagger}$\thanks{Corresponding author. Email: \protect\href{http://aryanfar@caltech.edu}{aryanfar@caltech.edu}.},
Dimitri M. Saad$^{\dagger}$ and William A. Goddard III$^{\S}$}
\maketitle

\lyxaddress{\begin{center}
\emph{$\dagger$ American University of Beirut, Riad ElSolh, Beirut,
Lebanon 1107-2020}\\
\emph{$\ddagger$ Bahçe\c{s}ehir University, 4 Ç\i ra\u{g}an Cad,
Be\c{s}ikta\c{s}, Istanbul, Turkey 34353}\\
\emph{$\S$ California Institute of Technology, 1200 E California
Blvd, Pasadena, CA 91125}
\par\end{center}}
\begin{abstract}
The amorphous propagation of microstructures during the electrochemical
charging of a battery is the main reason for the capacity decay and
short circuit. The charge distribution across the micro-structure
is the result of both local and global equilibrium and is non-convex
problem merely due to stochastic placement of the atoms. As such,
obtaining the charge equilibrium ($QEq$) is a critical factor, since
the amount of charge determines the success rate of the bond formation
for the ionic species approaching the microstructure and consequently
the ultimate morphology of the electrochemical deposits. Herein we
develop a computationally-affordable method for determining the charge
allocation within such microstructures. The cost function and the
span of the charge distribution correlates very closely with the trivial
method as well as a conventional method, albeit having significantly
less computational cost. The method can be used for optimization in
non-convex environments, specially those of stochastic nature. 
\end{abstract}

\section{Introduction}

\noindent Metallic anodes such as lithium, sodium and zinc are arguably
highly attractive candidates for use in high-energy and high-power
density rechargeable batteries \cite{Li14Review,Pei14,Slater13}.
In particular, lithium metal possess the lowest density and smallest
ionic radius which provides a very high gravimetric energy density
and possesses the highest electropositivity ($E^{0}=-3.04V$ vs SHE)
that likely provides the highest possible voltage, making it suitable
for high-power applications such as electric vehicles. ($\rho=0.53~g.cm^{-3}$)\cite{Li19Lithium,Xu14}.
During the charging, the fast-pace formation of microstructures with
relatively low surface energy from Brownian dynamics, leads to the
branched evolution with high surface to volume ratio \cite{Xu04}.
The quickening tree-like morphologies could occupy a large volume,
possibly reach the counter-electrode and short the cell (Figure \ref{fig:Naked}).
Additionally, they can also dissolve from their thinner necks during
subsequent discharge period. Such a formation-dissolution cycle is
particularly prominent for the metal electrodes due to lack of intercalation\footnote{Intercalation: diffusion into inner layer as the housing for the charge,
as opposed to depositing in the surface.}\cite{Li14Review}. Previous studies have investigated various factors
on dendritic formation such as current density\cite{Orsini98}, electrode
surface roughness \cite{Monroe04,Nielsen15,Natisiavas16}, impurities
\cite{Steiger14}, solvent and electrolyte chemical composition \cite{Schweikert13,Younesi15},
electrolyte concentration \cite{Brissot99Concentration}, utilization
of powder electrodes \cite{Seong08} and adhesive polymers\cite{Stone12},
temperature \cite{Aryanfar15Annealing}, guiding scaffolds \cite{Yao19,Qian19},
capillary pressure \cite{Deng19}, cathode morphology \cite{Abboud19}
and mechanics \cite{Xu17,Wang19}. Some of conventional characterization
techniques used include NMR \cite{Bhattacharyya10} and MRI \cite{Chandrashekar12}.
Recent studies also have shown the necessity of stability of solid
electrolyte interphase (i.e. SEI) layer for controlling the nucleation
and growth of the branched medium \cite{Li19Energy,Kasmaee16}. 

\noindent Earlier model of dendrites had focused on the electric field
and space charge as the main responsible mechanism \cite{Chazalviel90}
while the later models focused on ionic concentration causing the
diffusion limited aggregation (DLA) \cite{Monroe03,Witten83,Zhang19}.
Both mechanisms are part of the electrochemical potential \cite{Bard80,Tewari19},
indicating that each could be dominant depending on the localizations
of the electric potential or ionic concentration within the medium.
Nevertheless, their interplay has been explored rarely, especially
in continuum scale and realistic time intervals, matching scales of
the experimental time and space.

\noindent Dendrites instigation is rooted in the non-uniformity of
electrode surface morphology at the atomic scale combined with Brownian
ionic motion during electrodeposition. Any asperity in the surface
provides a sharp electric field that attracts the upcoming ions as
a deposition sink. Indeed the closeness of a convex surface to the
counter electrode, as the source of ionic release, is another contributing
factor. In fact, the same mechanism is responsible for the further
semi-exponential growth of dendrites in any scale. During each pulse
period the ions accumulate at the dendrites tips (unfavorable) due
to high electric field in convex geometry and during each subsequent
rest period the ions tend to diffuse away to other less concentrated
regions (favorable). The relaxation of ionic concentration during
the idle period provides a useful mechanism to achieve uniform deposition
and growth during the subsequent pulse interval. Such dynamics typically
occurs within the double layer (or stern layer \cite{Bazant11}) which
is relatively small and comparable to the Debye length. In high charge
rates, the ionic concentration is depleted and concentration on the
depletion reaches zero~\cite{Fleury97}; nonetheless, our continuum-level
study extends to larger scale, beyond the double layer region \cite{Aryanfar15Thermal}. 

\noindent Various charging protocols have been utilized for the prevention
of dendrites \cite{Li01}, which has previously been used for uniform
electroplating \cite{Chandrashekar08}. We have proven that the optimum
rest period for the suppression of dendrites correlates with the relaxation
time of the double layer for the blocking electrodes which is interpreted
as the \emph{RC time} of the electrochemical system \cite{Bazant04}.
We have explained qualitatively how relatively longer pulse periods
with identical duty cycles will lead to longer and more quickening
growing dendrites \cite{Aryanfar14Dynamics}. We developed coarse
grained computationally affordable algorithm that allowed us reach
to the experimental time scale (\emph{$\sim ms$}). Additionally,
in the recent theoretical work we indicated that there is an analytical
criterion for the optimal inhibition of growing dendrites \cite{Aryanfar18}. 

Additionally, the ultimate morphology of the dendritic electrodeposites,
depends on the possibility of the bond-formation when ion reaches
the outer boundary of the microstructure. The success of electron
transfer in such approach would highly be determined to the amount
of the charge presents in the electron transfer site. Therefore, in
this paper, we elaborate on the charge distribution in equilibrium
across the dendritic microstructures, where the placement of the stochastically-grown
dendrites. Subsequently, we verify our method via comparison with
trivial method, which is far more computationally expensive as well
as a conventional package. The convexification of charge distribution
can be utilized widely, as well as the charge equilibrium in the electrodeposits. 

\section{Methodology}

\subsection{Computational Method}
\noindent \begin{center}
\begin{figure}
\subfloat[Lab-scale observation of amorphous microstructures.\label{fig:Naked}
\cite{Aryanfar15Annealing}]{\raggedright{}\includegraphics[height=0.3\textheight]{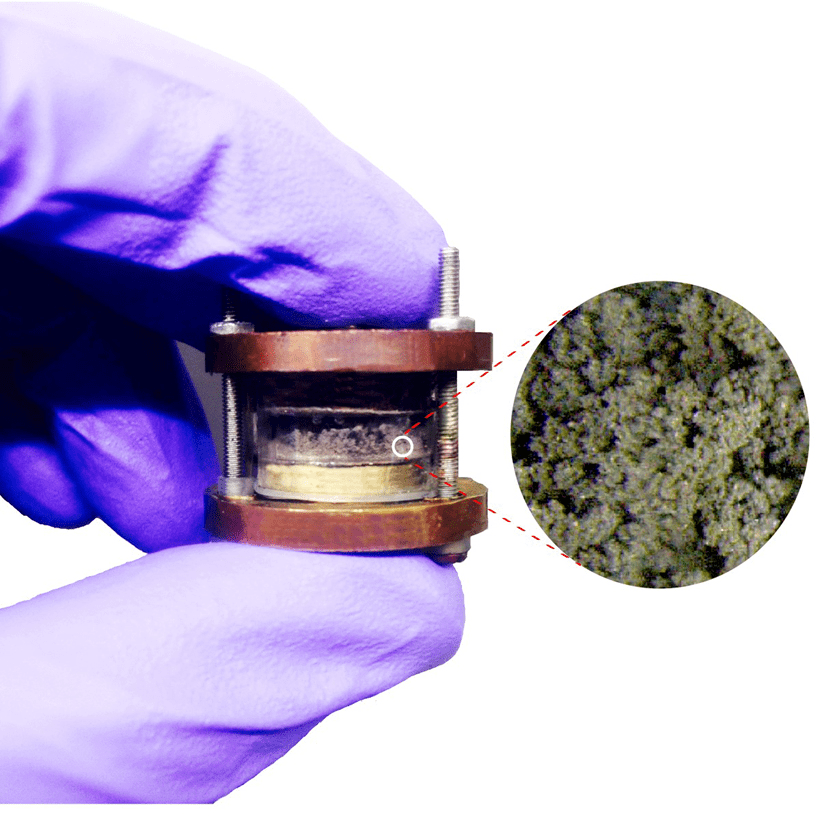}}\hfill{}\subfloat[The transport elements in the coarse scale of time, representing branch
vector $\vec{r}_{dend}$.\label{fig:Displacements} \cite{Aryanfar19Finite}]{\begin{raggedleft}
\includegraphics[height=0.3\textheight]{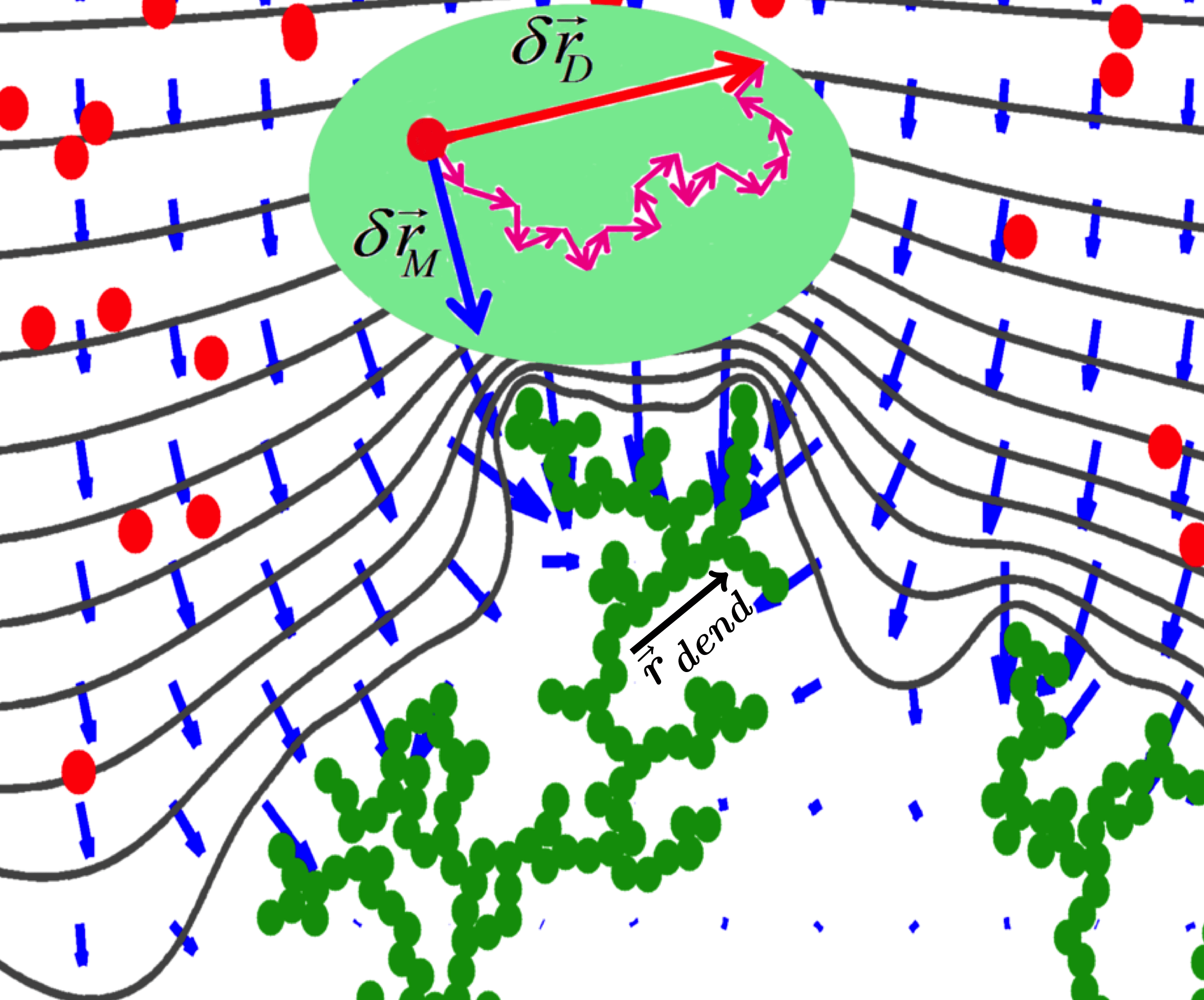}
\par\end{raggedleft}
}\caption{Observation/Modeling of dendritic propagation. \label{fig:Dendrites} }
\end{figure}
\par\end{center}

\noindent Figure \ref{fig:Dendrites} represents the dendritic propagation
in the lab scale as well as in our computations. The ionic flux is
generated in response to the variation of the electrochemical potential,
which is per see the result of the variations (i.e. gradient) of concentration
($\nabla C$) or electric potential ($\nabla V$). In the ionic scale,
the regions of higher concentration tend to collide and repel more
and, given enough time, diffuse to lower concentration zones, following
Brownian motion. Such inter-collisions could be added-up in the larger
scale and be addressed via diffusion length \cite{Aryanfar14Dynamics}
\footnote{\textcolor{black}{The diffusion coefficient $D^{+}$ is generally
concentration dependent \cite{Chandrashekar16}, due to dilute concentration
in the electro-neutral region, we assume the negligible variations.}} representing the average progress of a diffusive wave in a given
time and is obtained directly from the diffusion equation \cite{Philibert06}.
On the other hand, ions tend to acquire drift velocity in the electrolyte
medium when exposed to electric field and during the given time $\delta t$
their progress by the drift velocity. 

\noindent Therefore the total effective displacement $\delta\vec{\textbf{r}}$
with neglecting convection\footnote{Convection-wise, since the Rayleigh number $Ra$ is highly dependent
to the thickness (i.e. $Ra\propto l^{3}$), for a thin layer of electrodeposition
we have $Ra<1500$ and thus the convection is negligible. \cite{Fox16}} would be:

\noindent 
\begin{equation}
\vec{\textbf{r}}\left(t+\delta t\right)=\vec{\textbf{r}}\left(t\right)+\sqrt{2D^{+}\delta t}\text{ }\hat{\textbf{g}}+\mu^{+}\vec{\textbf{E}}\delta t\label{eq:TotalDis}
\end{equation}

\noindent where $D^{+}$ is the ionic diffusion coefficient in the
electrolyte, $\delta t$ is the coarse time interval\footnote{$\delta t=\sum_{i=1}^{n}\delta t_{i}$ where $\delta t_{k}$ is the
inter-collision time, typically in the range of $fs$.}, $\hat{\textbf{g}}$ is a normalized vector in random direction,
representing the Brownian dynamics, $\mu^{+}$ is the mobility of
cations in electrolyte and $\vec{\textbf{E}}$ is the local electric
field, which is the gradient of electric potential ($\vec{\textbf{E}}=-\nabla V$
). Such vector sum is represented in the Figure \ref{fig:Displacements}. 

\noindent The probability of successful jump depends on how much charge
each atom has from the sea of electrons. Such charge equilibrium would
be obtained from the minimization of the total potential energy for
the amorphous material, which is generally obtained by means of Taylor
expansion as \cite{Rappe91}: 

\noindent 
\begin{equation}
E_{A}(Q)=E_{A0}+Q_{A}\left(\frac{\partial E}{\partial Q}\right)_{A0}+\frac{1}{2}Q_{A}^{2}\left(\frac{\partial^{2}E}{\partial Q^{2}}\right)_{A0}+...\label{eq:E}
\end{equation}

\noindent where the second term in Equation \ref{eq:E} is in fact
the electronegativity $\chi$ and is defined by: 

\noindent 
\[
\chi_{A}=\frac{\partial E}{\partial Q_{A}}
\]

\noindent since all the composing atoms in dendrites are identical
the variation in fact breaks down to the cost function as the \emph{difference}
in the total potential energy $E(Q)$ based on the charge allocation
as: 

\noindent 
\begin{equation}
E(Q)={\displaystyle \sum_{i=1}^{n}\sum_{j=i+1}^{n}{\displaystyle \frac{q_{i}q_{j}}{d_{i,j}}}}\label{eq:Lambda}
\end{equation}

\noindent \begin{center}
\begin{figure}
\begin{centering}
\includegraphics[width=0.8\textwidth,height=0.35\textheight]{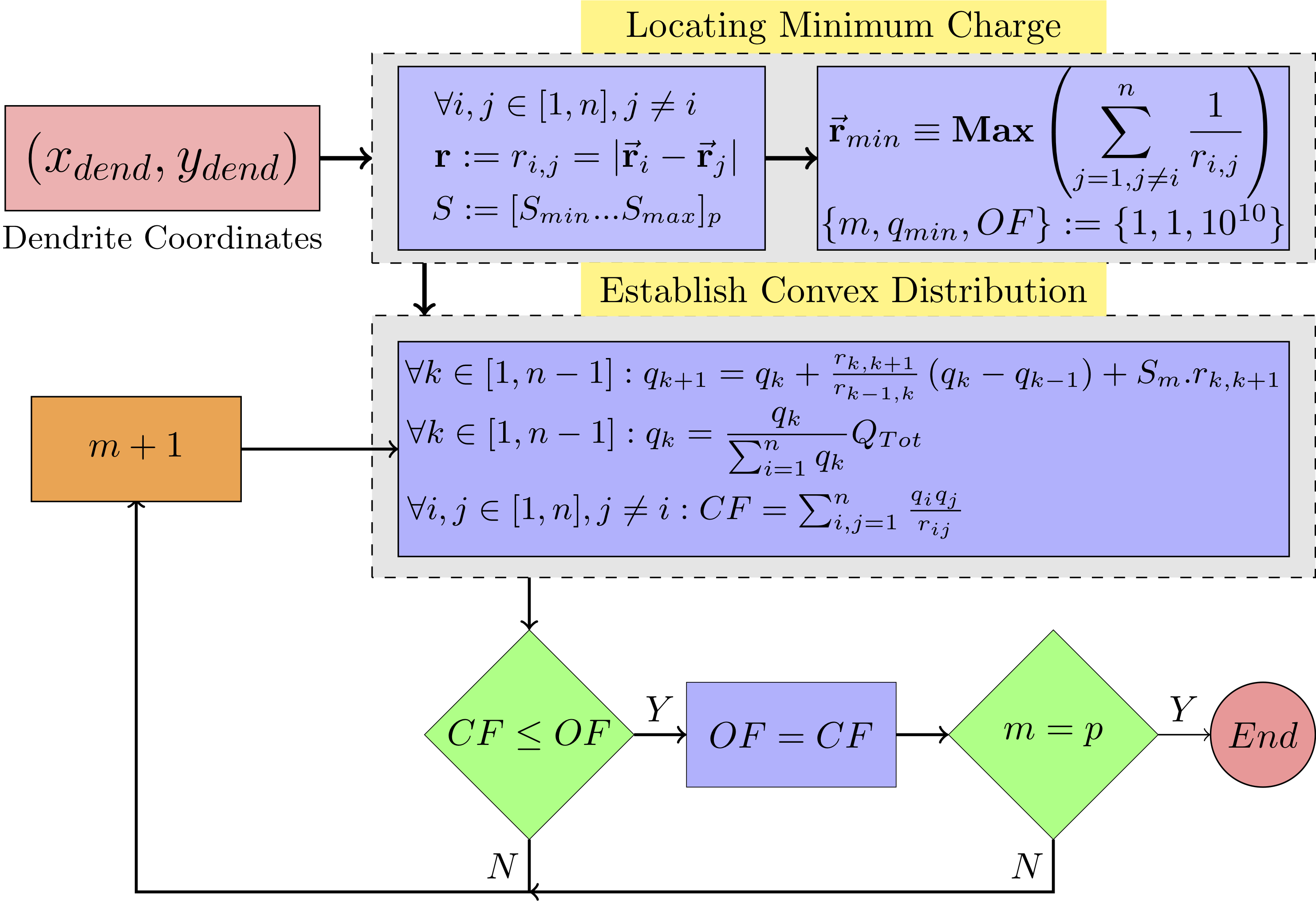}
\par\end{centering}
\caption{The convexification pseudo-code.\label{fig:Flowchart}}
\end{figure}
\par\end{center}

\noindent where $d_{i,j}$ is the interatomic distance from $i$ to
$j$ defined as: 

\noindent 
\[
d_{i,j}=|\vec{r}_{j}-\vec{r}_{i}|
\]

\noindent and $\vec{r}_{i}$ and $\vec{r}_{j}$ are the coordinates
of the atoms relative to a reference point. Assuming that the set
of charge values could be represented by the vector $q=[\begin{array}{ccc}
q_{1} & ... & q_{n}\end{array}]$ , one can interpret the optimization problem in the quadratic form
as: 

\noindent 
\[
\text{minimize }\frac{1}{2}q^{T}Rq
\]

\noindent \vspace{-0.75cm}

\noindent 
\begin{equation}
\begin{array}{cc}
\text{s.t.} & \begin{cases}
\sum_{i=1}^{n}q_{i}=Q\\
0\leq q_{i}\leq ne
\end{cases}\end{array}\label{eq:Cost}
\end{equation}
where the total energy is in fact the matrix from of Equation \ref{eq:Lambda}
and the reciprocal distance matrix $R$ is stablished: 

\noindent 
\[
R=\left[\begin{array}{cccc}
0 & {\displaystyle \frac{1}{d_{1,2}}} & \cdots & {\displaystyle \frac{1}{d_{1,n}}}\\
{\displaystyle \frac{1}{d_{2,1}}} & \ddots &  & \vdots\\
\vdots &  & \ddots\\
{\displaystyle \frac{1}{d_{n,1}}} & \cdots &  & 0
\end{array}\right]
\]

\noindent and $n$ and $e$ are the valence electrons and electron
charge respectively. The first constraint in Equation \ref{eq:Cost}
means that the total sum of charges is a constant value $Q$ given
to the dendrite and the second and third constraints determine the
capacity range of charge fraction for each atom in the microstructure.

\subsubsection{Locating minimum charge}

\noindent The energy difference defined by Equation \ref{eq:Cost}
as the cost function is depends on the charge allocation in the atoms
as well as their distance. Assuming the given charge $q_{i}$ to a
charge, in order to minimize the energy term ${\displaystyle \frac{q_{i}q_{j}}{2d_{i,j}}}$,
the corresponding charges $q_{j}$ with the lower distance (closer)
should contain lower relative charge values and vice versa. In fact
the allocation of charges should be such that the most populated atomic
regions in the crystal, formed by the random walk procedure, should
contain the lowest fraction of the charge, since the denominator in
the Equation \ref{eq:Lambda} is quite large in those regions. Such
position in fact could be obtained for the case of even distribution
in the atomic charges. Assuming $q_{min}$ as such position, the value
of the reciprocal distance sum ${\displaystyle \frac{1}{d_{min,k}}}$
should be the maximum: 

\noindent 
\begin{equation}
\underset{k}{max}\sum_{i=1}^{n-1}\frac{1}{d_{min,k}}\label{eq:Min}
\end{equation}

\noindent The minimum charge $q_{min}$ via this this expression signifies
it's the closest proximity to the other charges.
\noindent \begin{center}
\begin{figure}
\centering{}\includegraphics[height=0.3\textheight]{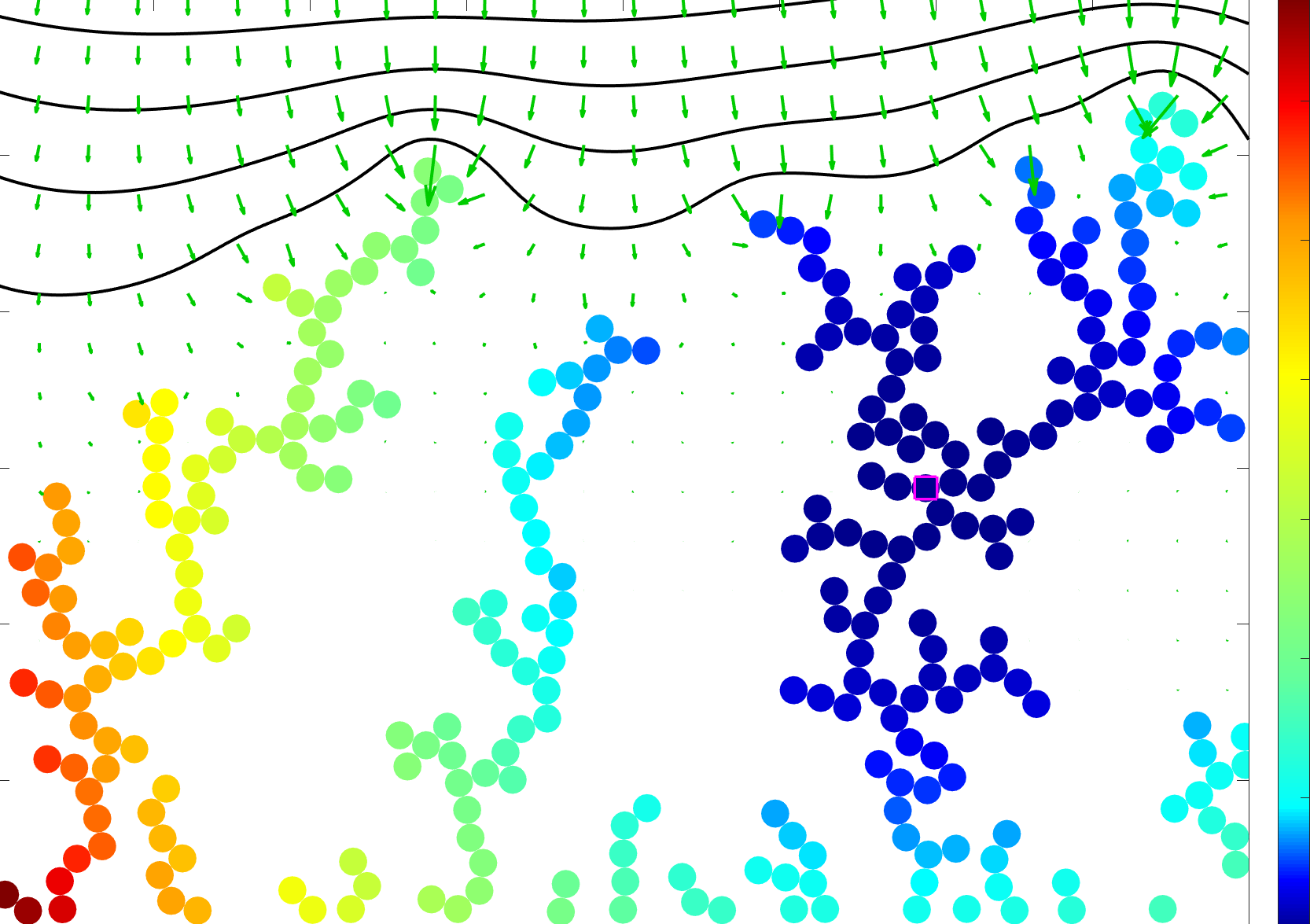}\caption{Dimension-less Charge distribution across the microstructure. The
minimum charge is illustrated with square.\label{fig:ChargeDistribution}}
\end{figure}
\par\end{center}

\subsubsection{Convexification}

\noindent \bigskip{}

\noindent Finding the closest proximity of the minimum charge $q_{min}$
in the Equation \ref{eq:Min} ensures the closest radial distance
to the surrounding atoms. In other words, the largest allocation of
charge magnitude should be given to the atoms farthest from the the
most compact regions. Therefore, starting from the minimum charge
$q_{min}$ as the reference and moving outward radially, any charge
distribution should have an increasing trend, and there will be no
sensitivity for variation in the azimuthal direction as: 

\noindent 
\[
\frac{\delta q}{\delta\theta}=0
\]

\noindent As well, the quadratic form of the potential energy $E(Q)$
in Equation \ref{eq:Lambda} suggests that the charge distribution
in the radial direction has convex shape, which we translate in the
radial direction $r$ to: 

\noindent 
\[
\frac{\delta^{2}q}{\delta r^{2}}\geq0
\]

\noindent Performing numerical segmentation, this typically leads
to $q_{i+1}-2q_{i}+q_{i-1}\ge0$ for consecutive charges. Since the
atoms possess non-uniform spacing, we arrive at: 

\noindent 
\[
{\displaystyle {\displaystyle {\displaystyle \frac{{\displaystyle \frac{\delta q_{k+1}}{\delta r}}-{\displaystyle \frac{\delta q_{k}}{\delta r}}}{d_{k+1,k-1}}}}\geq0}
\]

\noindent noting the difference in the slope $m$ we arrive the following:

\noindent 
\[
{\displaystyle \frac{q_{k+1}-q_{k}}{d_{k+1,k}}-\frac{{\displaystyle q}_{k}-q_{k-1}}{d_{k,k-1}}=m}
\]

\noindent therefore the value of charge $q_{k+1}$ consecutively can
be obtained as:

\noindent 
\begin{equation}
{\displaystyle q_{k+1}=q_{k}+\frac{d_{k+1,k}}{d_{k,k-1}}\left(q_{k}-q_{k-1}\right)+md_{k+1,k}}\label{eq:Segmented}
\end{equation}

\noindent \begin{center}
\begin{table}
\begin{centering}
\begin{tabular}{|c|c|c|c|}
\hline 
Parameter & Symbol & Value & Unit\tabularnewline
\hline 
\hline 
\# atoms & $n$ & $300$ & $[]$\tabularnewline
\hline 
Diffusivity & $D^{+}$ & $1.4\times10^{-14}$ & $m^{2}.s^{-1}$\tabularnewline
\hline 
Permittivity & $\varepsilon$ & $64$ & $[]$\tabularnewline
\hline 
Temperature & $T$ & $293$ & $K$\tabularnewline
\hline 
Domain length & $l$ & $180$ & $nm$\tabularnewline
\hline 
Voltage & $\Delta V$ & $0.1$ & $V$\tabularnewline
\hline 
\end{tabular}
\par\end{centering}
\caption{Parameters for sample computation \cite{Aryanfar14Dynamics}. \label{tab:Pars}}
\end{table}
\par\end{center}

\noindent The Equation \ref{eq:Segmented} assigns to a charge $i$
a value that ascribes convexity to the distribution given the charge
value of $2$ preceding points. Furthermore, it establishes convexification
of randomly distributed points starting from a minimum charged atom.
Such iteration has been performed for multiple values of the variation
difference $m$ for the charge distribution based on the constraints
in the Equation \ref{eq:Cost} such that the minimum value of the
objective function (Eq. \ref{eq:Lambda}) is obtained. The outlined
algorithm has been visualized in the Figure \ref{fig:Flowchart} based
on the parameters given in the Table \ref{tab:Pars}. Figure \ref{fig:ChargeDistribution}
shows the charge distribution within the given stochastically-developed
microstructure, where the location of the minimum charge is highlighted.
The black lines and green vectors represent the iso-potential contours
and electric field respectively. 

\subsection{Sample Computation}

\noindent We carry out the sample computation for the method we have
developed and we compare it against the conventional MATLAB framework,
as well as the trivial solutions, and filtered trivial solution. We
study the one dimensional case where atoms are allocated in a straight
line based on the numbers given in the Table \ref{tab:Verification}.
Here we propose the following methods to compare to: 
\noindent \begin{center}
\begin{table}
\begin{centering}
\begin{tabular}{|c|c|c|c|c|c|}
\hline 
$Q$ & $n$ & $x_{min}$ & $x_{max}$ & $q_{min}$ & $q_{max}$\tabularnewline
\hline 
\hline 
11 & 11 & -10 & 10 & 0 & 3.66\tabularnewline
\hline 
\end{tabular}
\par\end{centering}
\caption{Verification parameters. \label{tab:Verification}}
\end{table}
\par\end{center}

\subsubsection{Analytical Solution }

\noindent Since the central position from Equation \ref{eq:Lambda}
can be regarded as the minimum charge point, we can interpret that
the center of the 1D line could be the location for the minimum charge.
From the constraints in the Equation \ref{eq:Cost}, the type of analytical
function can be extracted. The charge should have the increasing slope
condition as well as positive second derivative for the formation
of the convex shape. If $x$ is the one dimensional coordinates, therefore
analogous to the given constraints the forms are obtained as: 

\noindent 
\[
\begin{cases}
{\displaystyle \frac{\partial\rho}{\partial x}\geq0} & \sim\alpha x\\
{\displaystyle \frac{\partial^{2}\rho}{\partial x^{2}}\geq0} & \sim\beta\exp(x)
\end{cases}
\]

\noindent where $\{\alpha,\beta\}>0$. The line will have a symmetric
charge distribution, one could study one of it's identical halves,
using the combinatorics the two forms via absorbing the two coefficients
$\alpha$ and $\beta$ into the new pre-factor $a$. Considering the
continuum-scale linear charge density $\rho(x)$ the charge distribution
would have a form of: 

\noindent 
\[
\rho(x)=ax\exp(bx)
\]

\noindent Therefore the energy minimization in Equation \ref{eq:Cost}
will translate into the following: 

\noindent 
\[
\text{minimize }\int_{0}^{l}\frac{\rho(x)}{x}dx
\]

\noindent \vspace{-0.75cm}

\noindent 
\begin{equation}
\begin{array}{cc}
\text{s.t.} & \begin{cases}
\int_{0}^{l}\rho(x)dx=Q\\
0\leq\rho(x)\leq ne
\end{cases}\end{array}\label{eq:MinimizeCont}
\end{equation}

\subsubsection*{Exact solution}

\noindent Assuming: ${\displaystyle \rho=axe^{bx}}$, one needs to
find two free parameters $a$ , $b$. The total energy $E(Q)$ can
be obtained as: 

\noindent 
\[
E(Q)=\frac{a}{b}{\displaystyle \left(\exp(bl)-1\right)}
\]

\noindent and the constraints will be obtained using the chain derivative
rule as: 

\noindent 
\begin{align*}
Q=\int_{0}^{l}ax\exp(bx)dx\\
=\frac{a}{b}x\exp(bx)\Bigg|_{0}^{l}-\int_{0}^{l}\frac{a}{b}\exp(bx)\\
=\left(\frac{al}{b}e^{bl}-\frac{a}{b^{2}}(e^{bl}-1)\right)
\end{align*}

\noindent Since the distribution form $\rho(x)$ is increasing the
boundary condition should satisfy at the end ($x=l$) and therefore: 

\noindent 
\[
\begin{cases}
{\displaystyle \frac{a}{{\displaystyle b}}}\left({\displaystyle le^{bl}}-{\displaystyle \frac{1}{b}e^{bl}}-{\displaystyle \frac{1}{b}})\right)=Q\\
ale^{bl}\leq:=1\\
a\geq0
\end{cases}
\]

\noindent Considering the maximum value in the inequality, we arrive
to the following via combination: 

\noindent 
\[
\left(\frac{1}{b}-\frac{1}{abl}-\frac{a}{b^{2}}\right)=Q
\]
This is a quadratic equation versus the exponent $b$. Therefore,
assuming $a=1$ it can be solved and the charge density is obtained
as: 

\noindent 
\begin{equation}
\rho(x)=x\exp\left(\left(\frac{l-1\pm\sqrt{(l-1)^{2}-4Ql}}{2Ql}\right)x\right)\label{eq:Exact}
\end{equation}

\subsubsection*{Simplified solution }

\noindent The exact form of charge distribution in Equation \ref{eq:Exact}
can be approximated with a simpler form. Here we prove that the exponent
has an upper bound of ${\displaystyle \frac{1}{l}}$: 

\noindent 
\[
\frac{l-1\pm\sqrt{(l-1)^{2}-4Ql}}{2Ql}\leq\frac{1}{l}
\]

\noindent Proving for maximum case one has: 

\noindent 
\begin{equation}
\sqrt{(l-1)^{2}-4Ql}\leq(1-l)+2Q\label{eq:Ineq}
\end{equation}

\noindent Since the $LHS$ is positive value, the $RHS$ we must have:
\[
l\leq2Q+1
\]

\noindent As well the square sign should be non-negative, therefore: 

\noindent 
\[
(l-1)^{2}-4Ql\geq0
\]

\noindent 
\[
l\leq2Q+1-2\sqrt{Q(Q+1)}
\]

\noindent Taking the inequality \ref{eq:Ineq} to the power 2 we get:

\noindent 
\[
0\leq Q(Q+1)\text{ \ensuremath{\checkmark}}
\]

\noindent which is always true and the upper bound is determined.
Therefore the exponent could be considered as $b:={\displaystyle \frac{1}{l}}$
. Thus the sum condition gives: 

\noindent 
\begin{align*}
Q & =\int_{0}^{l}a{\displaystyle xe^{{\displaystyle \frac{x}{l}}}=\left(alxe^{{\displaystyle \frac{x}{l}}}-\int_{0}^{l}ale^{{\displaystyle \frac{x}{l}}}\right)\Bigg|_{0}^{l}=al^{2}}
\end{align*}

\noindent and the coefficient $a$ is updated accordingly versus the
total charge sum $Q$. Hence, the simplified charge would be: 

\noindent 
\begin{equation}
\rho(x)={\displaystyle \frac{Q}{l^{2}}xe^{{\displaystyle \frac{x}{l}}}}\label{eq:Anal}
\end{equation}

Note that this is valid for $x\geq0$ and the other (negative) half
can be established from symmetry. 

\subsubsection{Enhanced Trivial Method}

\noindent Using the trivial method, we iteratively scan the charge
distribution $\{q_{1},...,q_{n}\}$ which minimizes the total energy
$E(Q)$ in Equation \ref{eq:Cost} from the all possible permutations
from combinatorics. The approximate iterative solution could be finding
the non-zero integer distribution to the pre-determined sum constraint
$Q$ given as: 

\noindent 
\begin{equation}
q_{1}+...+q_{n}=pQ\label{eq:Sum}
\end{equation}

\noindent where, due to integer nature of the solution, the fixed
total sum has been augmented $p$-fold to allow higher precision of
the distribution and the resulted distribution will ultimately scaled
back $p$-fold to satisfy the sum constraint in Equation \ref{eq:Cost}.
As well, the range constraints could been translated into the followings
to save a significant portion of the trivial solutions from Equation
\ref{eq:Sum}: 

\noindent 
\begin{equation}
\begin{cases}
q_{k+1}\geq q_{k}\\
q_{k+1}-2q_{k}+q_{k-1}\geq0
\end{cases}\label{eq:ConstraintsDiscrete}
\end{equation}

\noindent The next comparison has been performed with the conventional
MATLAB package function $fmincon$ in terms of cost and accuracy.
Since this function locates the local minima, it was run for $5$
different initial distribution values in the close proximity of the
analytical solution, to target the global minimum. As well, the initial
condition was given based on the analytical solution in the Equation
\ref{eq:Anal}. 

\noindent The resulted distributions have been visualized in the Figure
\ref{fig:Visual} and the corresponding significant numbers are compared
in the Figure \ref{fig:Visual}. It is worth noting that the computational
time required for the trivial case post-filtration is significantly
less than the filtered cases via Equation \ref{eq:ConstraintsDiscrete},
with the factor of $\sim10^{5}$.
\noindent \begin{center}
\begin{figure}
\subfloat[Visual. \label{fig:Visual}]{\includegraphics[height=0.3\textheight]{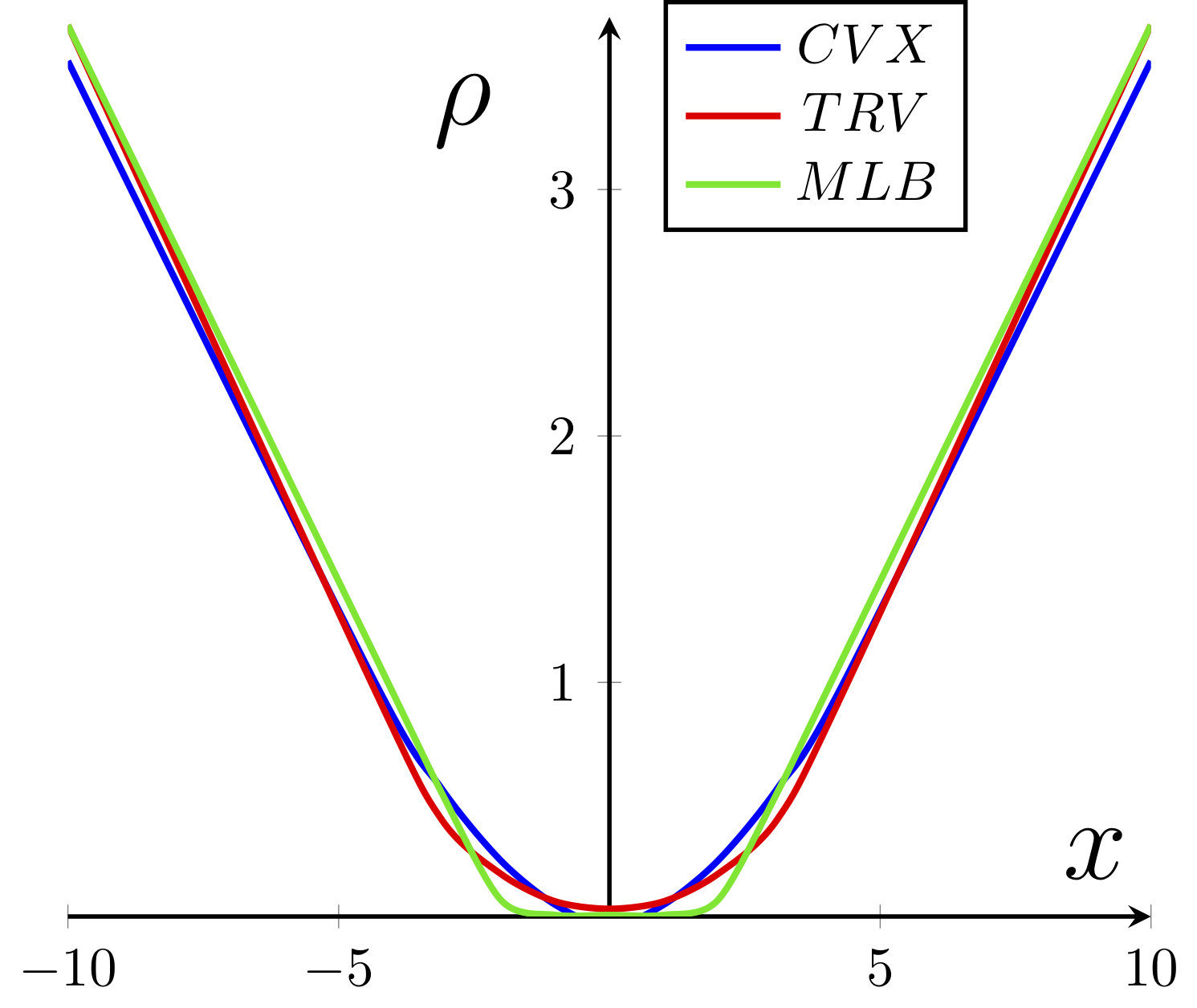}

}\subfloat[Numerical. \label{fig:Bars}]{\raggedleft{}\includegraphics[height=0.3\textheight]{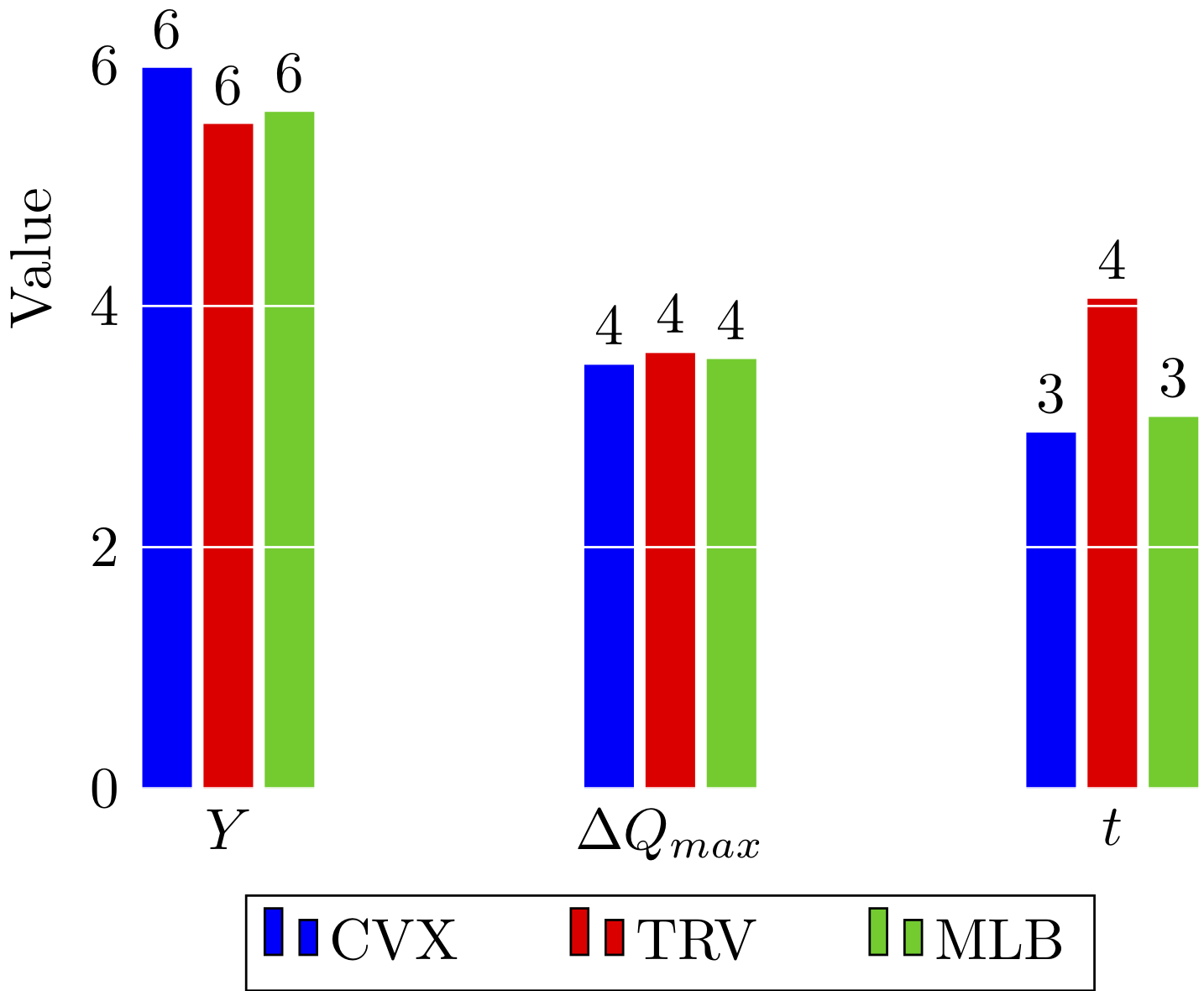}}

\caption{Comparison. \label{fig:Verification}}
\end{figure}
\par\end{center}

\section{Results \& Discussions}

\noindent Finding the minimum energy $E(Q)$ for amorphous microstructures
is usually a non-convex problem, which makes it difficult to solve.
This is merely due to stochastic allocation of the atoms. Since the
reciprocal distance matrix $R$ is symmetric ($R_{i,j}=R_{j,i}$),
for any given matrix $B$ there exists matrix $Z$ such that: 

\noindent 
\[
R=B^{-1}ZB
\]

\noindent Due to symmetry, the trace of reciprocal matrix $tr(R)$
is the sum of it's eigenvalues $\lambda_{i}$. Since, the distance
of each atom to itself is zero, $tr(R)=0$ and hence: 

\noindent 
\[
tr(R)=\sum\lambda_{i}=diag(R)=0
\]
the zero-sum constraint shows that at least one eigenvalue is negative.
Therefore the problem is not convex. 

The flowchart \ref{fig:Flowchart} shows the information flow for
determining the charge allocation leading to the minimum energy $E(Q)$,
which is mainly divided into two compartment of \emph{locating the
minimum charge}, and \emph{establishing a convex charge distribution}.
Such division in fact is an approximation, providing a significantly
less computational cost versus the \emph{whole-in }minimization of
the total energy $E(Q)$ given in the Equation \ref{eq:Lambda}. In
fact, the minimization of the reciprocal distance sum in the Equation
\ref{eq:Min} ensures that the most populated central (body) regions
would be the location of the minimum charge and vice versa, the outer
(boundary) regions would possess the highest portion of the charge
sum, since they will be farthest from the rest of the atoms to create
large potentials. This has been illustrated in the random dendritic
microstructure illustrated in the Figure \ref{fig:ChargeDistribution}.
Such allocation means that the outer atoms in fact will have higher
possibility for the electron donation rather than the inner layers. 

\noindent The comparative analysis of our method has been performed
for the 1D arrangement of the atoms based on the Table \ref{tab:Verification}.
The verification has been illustrated versus the enhanced trivial
search method as well as the commercial package. The accuracy comparison
is shown in the Figure \ref{fig:Visual} and the effectiveness is
represented in the Figure \ref{fig:Bars}. 
\noindent \begin{center}
\begin{figure}
\begin{centering}
\subfloat[Perturbations in convexity. \label{fig:Perturbation}]{\includegraphics[height=0.2\textheight]{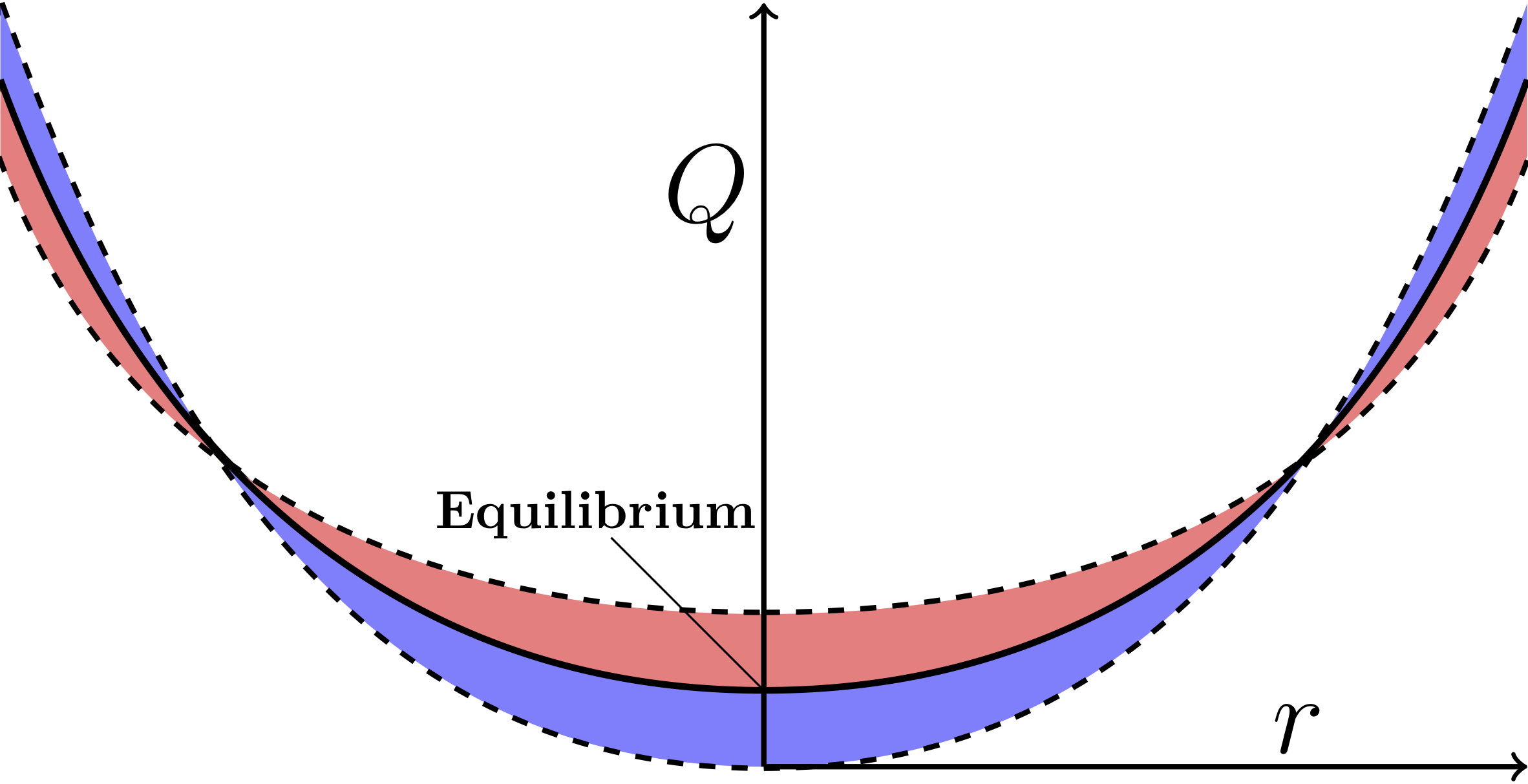}}\hfill{}\subfloat[Convex Hull. \label{fig:ConvexHull}]{\centering{}\includegraphics[height=0.2\textheight]{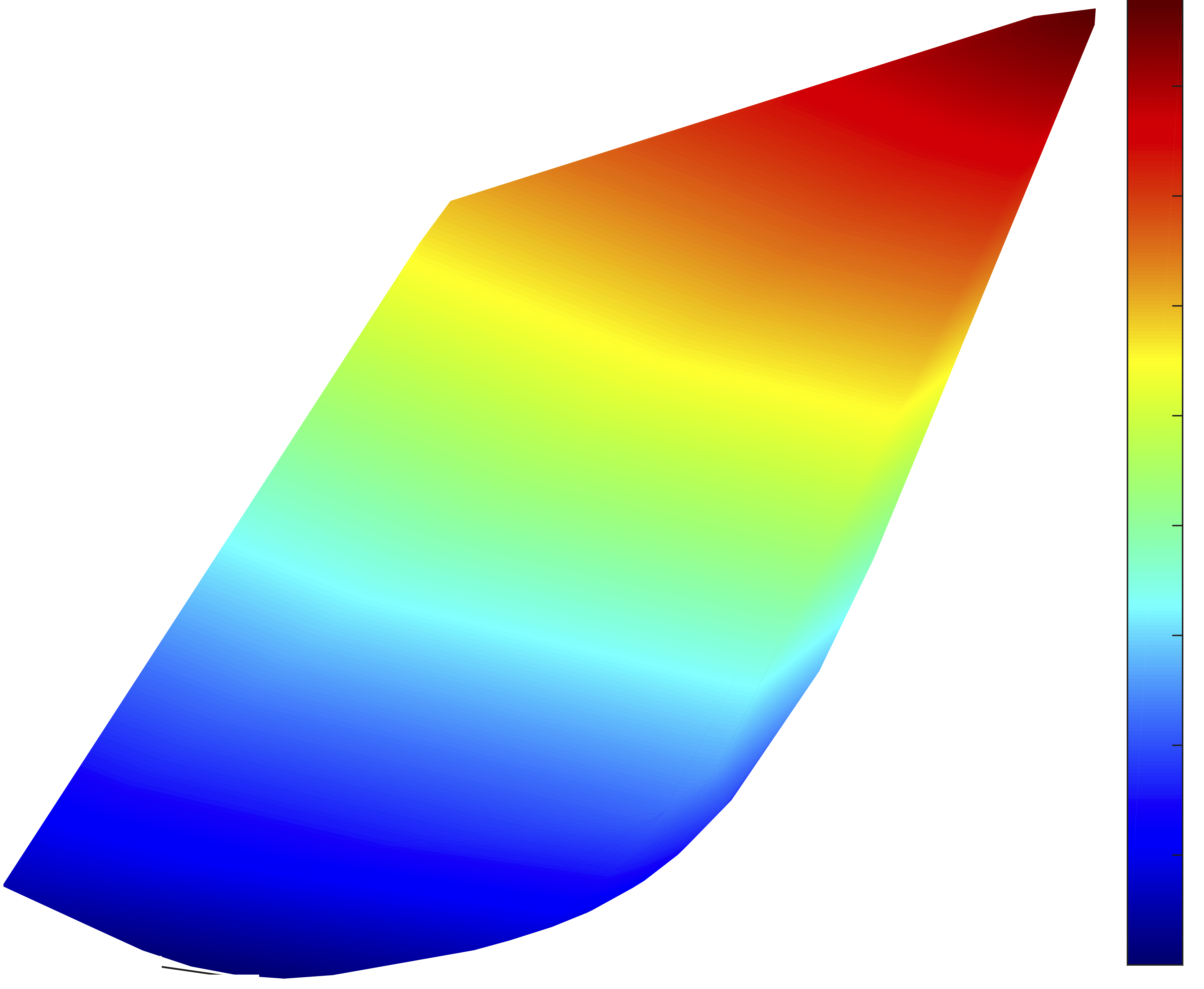}}
\par\end{centering}
\caption{Analysis.}

\end{figure}
\par\end{center}

Additionally the advantage of the developed method is that the result
is independent of the initial condition, as opposed to the commercial
package. Typical methods of finding the optimum solution requires
wither restricting the search within a convex set \cite{Acikmese13},
or relaxing non-convex constraints \cite{Liu17,Liu18}. Our method
is numerically following the same track, searching iteratively via
small perturbations in the convexity (Fig. \ref{fig:Perturbation})
to obtain the minimum energy $E(Q)$. Additionally the developed method
finds the optimum charge distribution via the physical and spatial
awareness of the convex dispersal of the charge. Such convex hull
is illustrated in the Figure \ref{fig:ConvexHull}, where the colors
qualitatively represent the relative charge values. 

\section{Conclusions}

\noindent In this paper, we developed a convexification method for
the charge equilibrium within the given stochastically-evolved dendritic
microstructure. Our computationally affordable method, which has mainly
been divided to simpler compartments, has been compared against the
conventional method as well as the commercial packages.

The significance of the method is the independence from the initial
condition and very low computational cost. Our method could be used
for determining the charge allocation in a larger clusters of microstructures
where the convex optimization would not be possible. In particular
the charge magnitude would determine the reaction probability, the
rate of propagation and the densification of microstructure during
the branched evolution. 

\section*{Acknowledgement}

\noindent We would like to thank internal support from the Faculty
of Engineering and Architecture at American University of Beirut. 

\bibliographystyle{unsrt}
\bibliography{Refs}

\end{document}